\documentclass[journal]{IEEEtran}

\ifCLASSINFOpdf
\else
   \usepackage[dvips]{graphicx}
\fi
\usepackage{url,amsfonts,amsmath,graphicx,enumitem,cite, hyperref, xcolor}
\usepackage{graphicx}

\begin{document}

\title{WaveCRN: An Efficient Convolutional Recurrent Neural Network for End-to-end Speech Enhancement}

\author{Tsun-An Hsieh, Hsin-Min Wang, Xugang Lu, and Yu Tsao, \IEEEmembership{Member, IEEE}\thanks{Thanks to MOST of Taiwan for funding.}}
\markboth{SPL-29109-2020.R1}
{Shell \MakeLowercase{{\em et al.}}: Bare Demo of IEEEtran.cls for IEEE Journals}
\maketitle
\setlength{\parskip}{0em}
\begin{abstract}
Due to the simple design pipeline, end-to-end (E2E) neural models for speech enhancement (SE) have attracted great interest. In order to improve the performance of the E2E model, the local and sequential properties of speech should be efficiently taken into account when modelling. However, in most current E2E models for SE, these properties are either not fully considered or are too complex to be realized. In this paper, we propose an efficient E2E SE model, termed WaveCRN. \textcolor{black}{Compared with models based on convolutional neural networks (CNN) or long short-term memory (LSTM), WaveCRN uses a CNN module to capture the speech locality features and a stacked simple recurrent units (SRU) module to model the sequential property of the locality features. Different from conventional recurrent neural networks and LSTM, SRU can be efficiently parallelized in calculation, with even fewer model parameters.} In order to more effectively suppress noise components in the noisy speech, we derive a novel restricted feature masking approach, which performs enhancement on the feature maps in the hidden layers; this is different from the approaches that apply the estimated ratio mask to the noisy spectral features, which is commonly used in speech separation methods. Experimental results on speech denoising and compressed speech restoration tasks confirm that with the SRU and the restricted feature map, WaveCRN performs comparably to other state-of-the-art approaches with notably reduced model complexity and inference time.
\end{abstract}


\IEEEpeerreviewmaketitle

\section{Introduction}
\label{sec:intro}
Speech related applications, such as automatic speech recognition (ASR), voice communication, and assistive hearing devices, play an important role in modern society. However, most of these applications are not robust when noises are involved. Therefore, speech enhancement (SE) \cite{SE, SETAN, DDAE1, WeinerDDAE, SUPERVISEDSS, meng2018adversarial, soni2018time, chai2019using}, which aims to improve the quality and intelligibility of the original speech signal, has been widely used in these applications.\par
In recent years, deep learning algorithms have been widely used to build SE systems. A class of SE systems carry out enhancement on the frequency-domain acoustic features, which is generally called spectral-mapping-based SE approaches. In these approaches, speech signals are analyzed and reconstructed using the short-time Fourier transform (STFT) and inverse STFT, respectively \cite{SE2, MLP1, wang2017transfer, liu2014experiments, Sun2017Multiple-target}. \textcolor{black}{Then, the deep learning models, such as fully connected deep denoising auto-encoder \cite{DDAE1}, convolutional neural networks (CNNs) \cite{SNRAWARE}, and recurrent neural networks (RNNs) and long short-term memory (LSTM) \cite{LSTM1, MaasLOVNN12}, are used as a transformation function to convert noisy spectral features to clean ones. In the meanwhile, some approaches are derived by combining different types of deep learning models (e.g., CNN and RNN) to more effectively capture the local and sequential correlations \cite{CRNSE, tan2019learning, CRN1, CRN2}. More recently, a SE system that was built based on stacked simple recurrent units (SRUs) \cite{sru, sruse} has shown denoising performance comparable to that of the LSTM-based SE system, while requiring much less computational costs for training.} Although the above-mentioned approaches can already provide outstanding performance, the enhanced speech signal cannot reach its perfection owing to the lack of accurate phase information. \textcolor{black} {To tackle this problem, some SE approaches adopt complex-ratio-masking and complex-spectral-mapping to enhance distorted speech \cite{fu2017complex, phase2, yao2019coarse}. In \cite{phasenet}, the phase estimation was formulated as a classification problem and was used in a source separation task.}\par
\textcolor{black}{Another class of SE methods proposes to directly perform enhancement on the raw waveform \cite{FCN1, rawwave1, TCNN, li2019single, kolbaek2020loss}, which are generally called waveform-mapping-based approaches. Among the deep learning models, fully convolutional networks (FCNs) have been widely used to directly perform waveform mapping \cite{ rawwave1,FCN2,SEGAN, Qian2017Speech}. The WaveNet model, which was originally proposed for text-to-speech tasks, was also used in the waveform-mapping-based SE systems \cite{Rethage2017Wavenet, waveunet}. Compared to a fully connected architecture, fully convolution layers retain better local information, and thus can more accurately model the frequency characteristics of speech waveforms. More recently, a temporal convolutional neural network (TCNN) \cite{TCNN} was proposed to accurately model temporal features and perform SE in the time domain. In addition to the point-to-point loss (such as $l_1$ and $l_2$ norms) for optimization, some waveform-mapping based SE approaches \cite{pascual2019time, DFL} utilized adversarial loss or perceptual loss to capture high-level distinctions between predictions and their targets.}\par
\textcolor{black}{For the above waveform-mapping-based SE approaches, an effective characterization of sequential and local patterns is an important consideration for the final SE performance. Although the combination of CNN and RNN/LSTM may be a feasible solution, the computational cost and model size of RNN/LSTM are high, which may considerably limit its applicability. In this study, we propose an E2E waveform-mapping-based SE method using an alternative CRN, termed WaveCRN\footnote{The implementation of WaveCRN is available at \url{https://github.com/aleXiehta/WaveCRN}.}, which combines the advantages of CNN and SRU to attain improved efficiency.} As compared to spectral-mapping-based CRN \cite{CRNSE, tan2019learning, CRN1, CRN2}, the proposed WaveCRN directly estimates feature masks from unprocessed waveforms through highly parallelizable recurrent units. Two tasks are used to test the proposed WaveCRN approach: (1) speech denoising and (2) compressed speech restoration. For speech denoising, we evaluate our method using an open-source dataset \cite{NDTSEA} and obtain high perceptual evaluation of speech quality (PESQ) scores \cite{PESQ} which is comparable to the state-of-the-art method while using a relatively simple architecture and $l_1$ loss function. \textcolor{black}{For compressed speech restoration, unlike in \cite{compress0, compress1} that used acoustic features, we simply pass the speech to a sign function for compression.} This task is evaluated on the TIMIT database \cite{TIMIT}. The proposed WaveCRN model recovers extremely compressed speech with a notable relative short-time objective intelligibility (STOI) \cite{STOI} improvement of 75.51\% (from 0.49 to 0.86).
\begin{figure}[t]
\begin{minipage}[b]{\linewidth}
\centering
\centerline{\includegraphics[width=\textwidth]{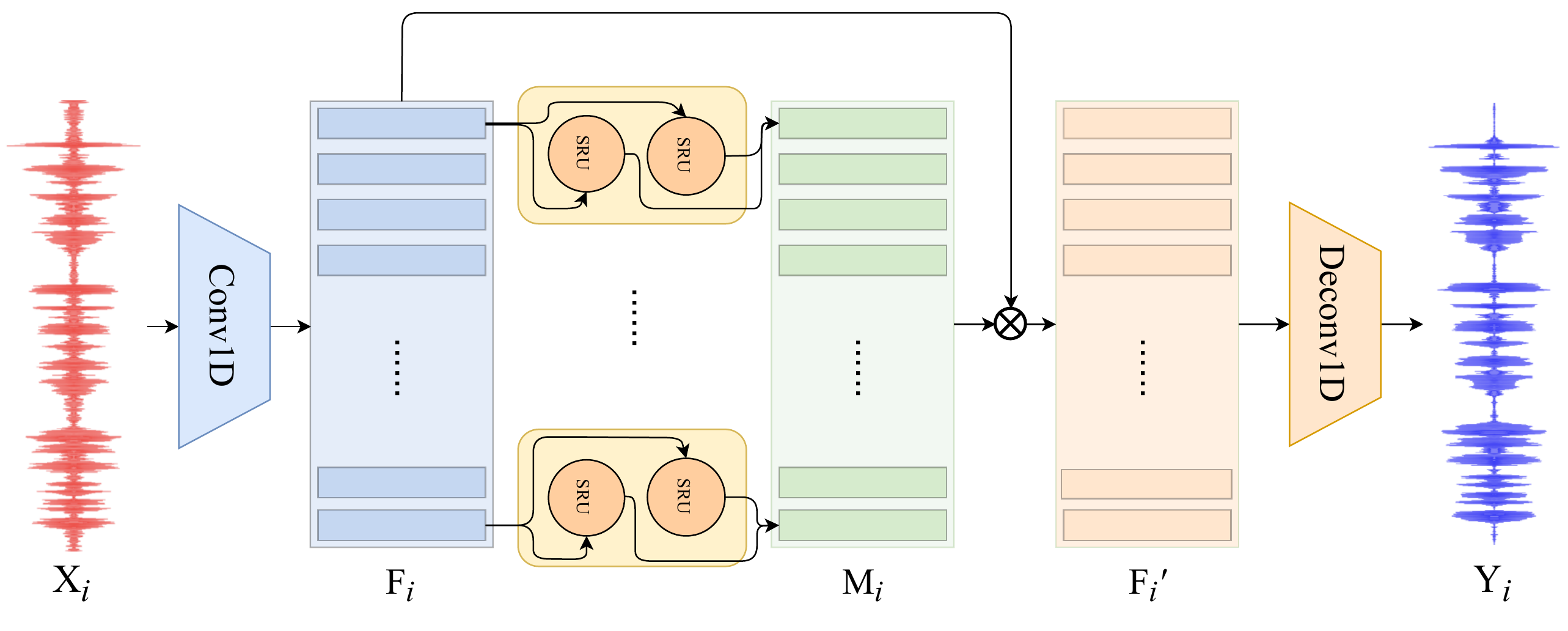}}
%
\end{minipage}
\caption{The architecture of the proposed WaveCRN model. Unlike spectral CRN, WaveCRN integrates 1D CNN with bidirectional SRU.}
\vspace{-0.5cm}
\label{fig:model}
\end{figure}

\section{Methodology}
\label{sec:part2}
In this section, we describe the details of our WaveCRN-based SE system. The architecture is a fully differentiable E2E neural network that does not require pre-processing and handcrafted features. Benefiting from the advantages of CNN and SRU, it jointly models local and sequential information. The overall architecture of WaveCRN is shown in Fig. \ref{fig:model}.
\subsection{1D Convolutional Input Module}
\label{ssec:WaveCRN-input}
As mentioned in the previous section, for the spectral-mapping-based SE approaches, speech waveforms are first converted to spectral-domain by STFT. To implement waveform-mapping SE, WaveCRN uses a 1D CNN input module to replace the STFT processing. Benefiting from the nature of neural networks, the CNN module is fully trainable. For each batch, the input noisy audio \textbf{X} (\textbf{X} $\in R^{N \times 1 \times L}$) is convolved with a two-dimensional tensor \textbf{W} (\textbf{W} $\in R^{C \times K}$) to extract the feature map \textbf{F} $\in R^{N \times C \times T}$, where $N, C, K,T, L$ are the batch size, number of channels, kernel size, time steps, and audio length, respectively. Notably, to reduce the sequence length for computational efficiency, we set the convolution stride to half the size of the kernel, so that the length of \textbf{F} is reduced from $L$ to $T = 2L/K + 1$.
\subsection{Temporal Encoder}
\label{ssec:subhead}
We used a bidirectional SRU (Bi-SRU) to capture the temporal correlation of the feature maps extracted by the input module in both directions. \textcolor{black}{For each batch, the feature map $\mathbf{F} \in R^{N \times C \times T}$ is passed to the SRU-based recurrent feature extractor. The hidden states extracted in both directions are concatenated to form the encoded features.}

\subsection{Restricted Feature Mask}
\label{ssec:subhead}
The optimal ratio mask (ORM) has been widely used in SE and speech separation tasks \cite{ORM}. As ORM is a time-frequency mask, it cannot be directly applied to waveform-mapping-based SE approaches. \textcolor{black}{In this study, an alternative mask restricted feature mask (RFM) with all elements in the range of -1 to 1 is applied to mask the feature map \textbf{F}:}
\begin{equation}
    \textbf{F}' = \textbf{M} \circ \textbf{F}.
\end{equation}
where $\textbf{M} \in R^{N \times C \times T}$, is the RFM, $\textbf{F}'$ is the masked feature map estimated by element-wise multiplying the mask \textbf{M} and the feature map \textbf{F}. It should be noted that the main difference between ORM and RFM is that the former is applied to spectral features, whereas the latter is used to transform the feature maps.
\subsection{Waveform Generation}
\label{ssec:subhead}
As described in Section \ref{ssec:WaveCRN-input}, the sequence length is reduced from $L$ (for the waveform) to $T$ (for the feature map) due to the stride in the convolution process. Length restoration is essential for generating an output waveform with the same length as the input. Given the input length, output length, stride, and padding as $L_{in}$, $L_{out}$, $S$, and $P$, the relationship between $L_{in}$ and $L_{out}$ can be formulated as:
\begin{gather}
    L_{out} = (L_{in} - 1) \times S - 2 \times P + (K - 1) + 1.
\end{gather}
Let $L_{in}=T$, $S=K/2$, $P=K/2$, we have $L_{out}=L$. That is, the output waveform and the input waveform are guaranteed to have the same length.
\subsection{Model Structure Overview}
\label{ssec:subhead}
\textcolor{black}{As shown in Fig. \ref{fig:model}, our model leverages the benefits of CNN and SRU. Given the $i$-th noisy speech utterance $\mathbf{X}_i \in R^{1 \times L}$, $i = 0, ..., N-1$, in a batch, a 1D CNN first maps $\mathbf{X}_i$ into a feature map $\mathbf{F}_i$ for local feature extraction. Bi-SRU then computes an RFM $\mathbf{M}_i$, which element-wisely multiplies $\mathbf{F}_i$ to generate a masked feature map $\mathbf{F}_i'$. Finally, a transposed 1D convolution layer recovers the enhanced speech waveforms, $\mathbf{X}_i$, from the masked features, $\mathbf{F}_i'$.}

\begin{figure}[t]
\begin{minipage}[b]{0.49\linewidth}
  \centering
  \centerline{\includegraphics[width=\textwidth]{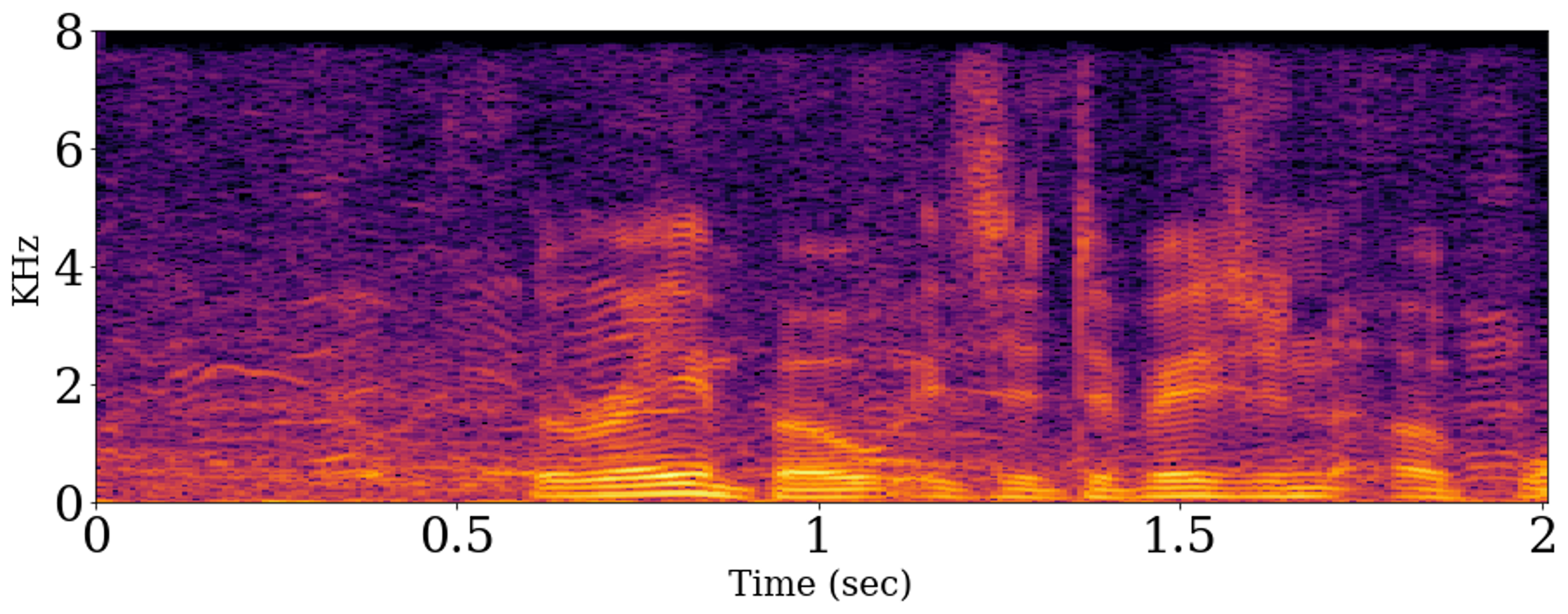}}
  \vspace{-0.1cm}
  \centerline{(a) Noisy}\medskip
\end{minipage}
\begin{minipage}[b]{0.49\linewidth}
  \centering
  \centerline{\includegraphics[width=\textwidth]{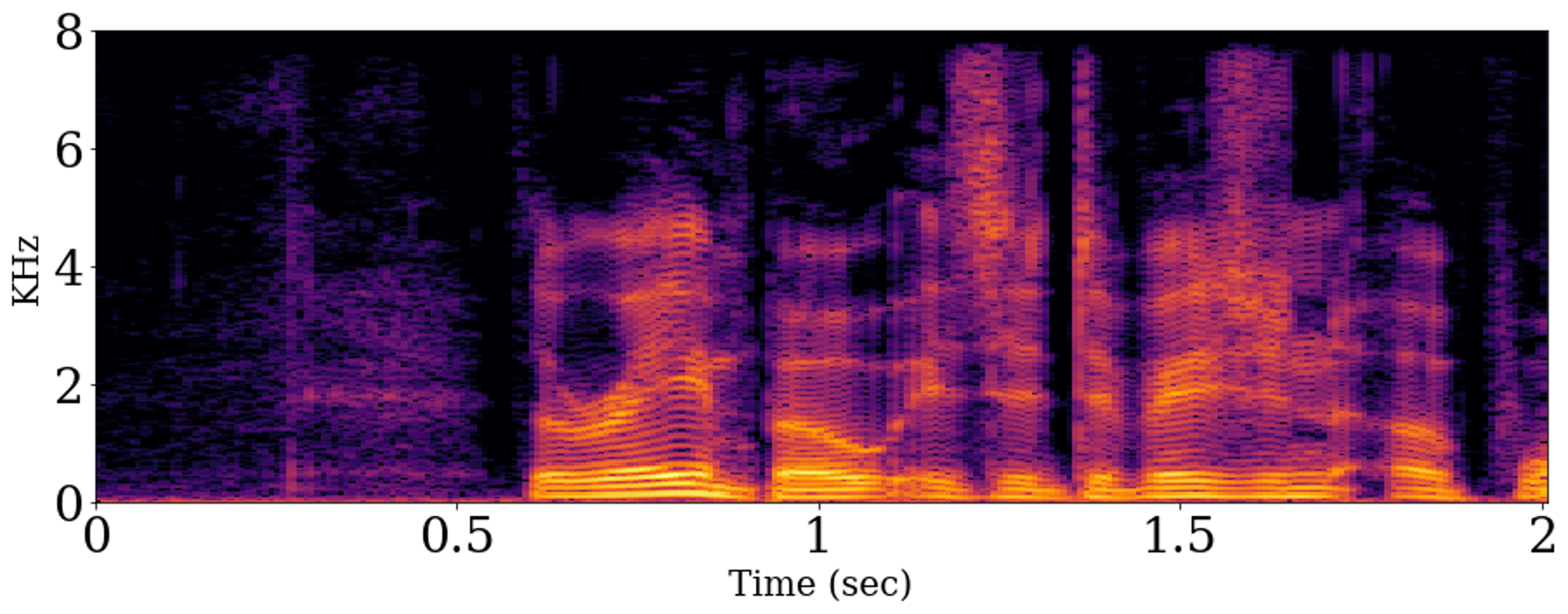}}
  \vspace{-0.1cm}
  \centerline{(b) Clean}\medskip
\end{minipage}
\begin{minipage}[b]{0.49\linewidth}
  \centering
  \vspace{-0.2cm}
  \centerline{\includegraphics[width=\textwidth]{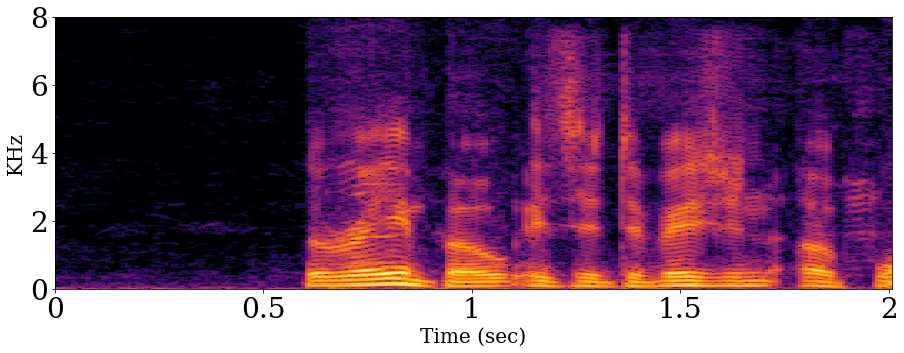}}
  \vspace{-0.1cm}
  \centerline{(c) LPS--SRU*}\medskip
\end{minipage}
\hfill
\begin{minipage}[b]{0.49\linewidth}
  \centering
  \vspace{-0.2cm}
  \centerline{\includegraphics[width=\textwidth]{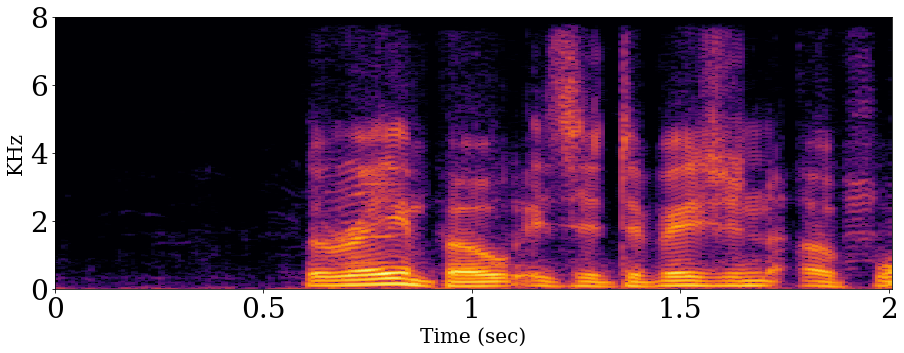}}
  \vspace{-0.1cm}
  \centerline{(d) LPS--SRU}\medskip
\end{minipage}
\begin{minipage}[b]{0.49\linewidth}
  \centering
  \vspace{-0.2cm}
  \centerline{\includegraphics[width=\textwidth]{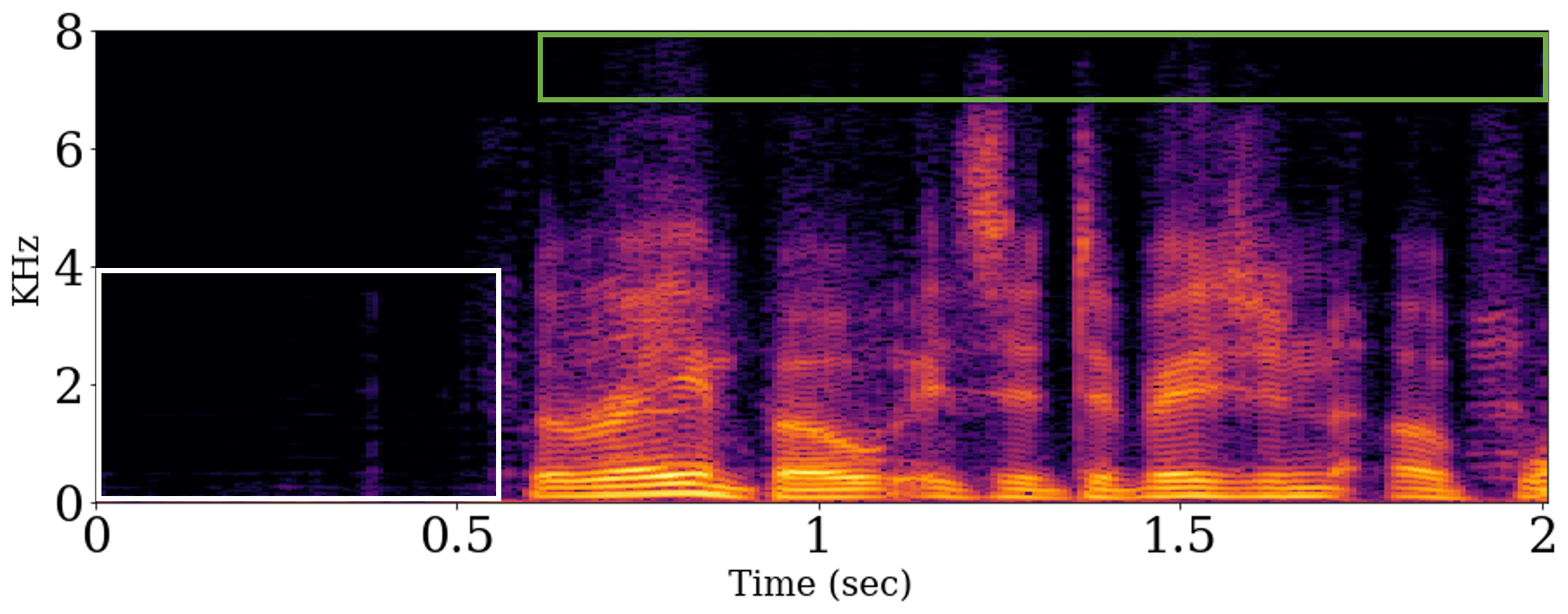}}
  \vspace{-0.1cm}
  \centerline{(e) WaveCBLSTM*}\medskip
\end{minipage}
\hfill
\begin{minipage}[b]{0.49\linewidth}
  \centering
  \vspace{-0.2cm}
  \centerline{\includegraphics[width=\textwidth]{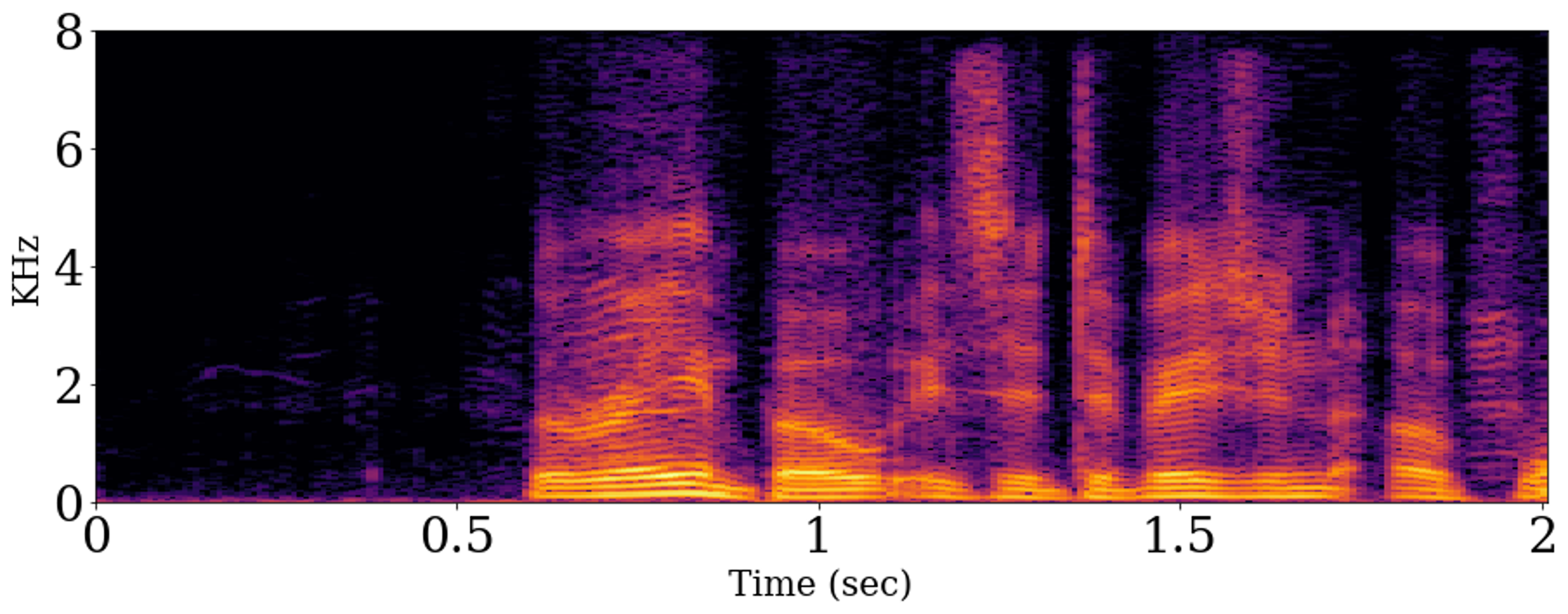}}
  \vspace{-0.1cm}
  \centerline{(f) WaveCBLSTM}\medskip
\end{minipage}
\begin{minipage}[b]{0.49\linewidth}
  \centering
  \vspace{-0.2cm}
  \centerline{\includegraphics[width=\textwidth]{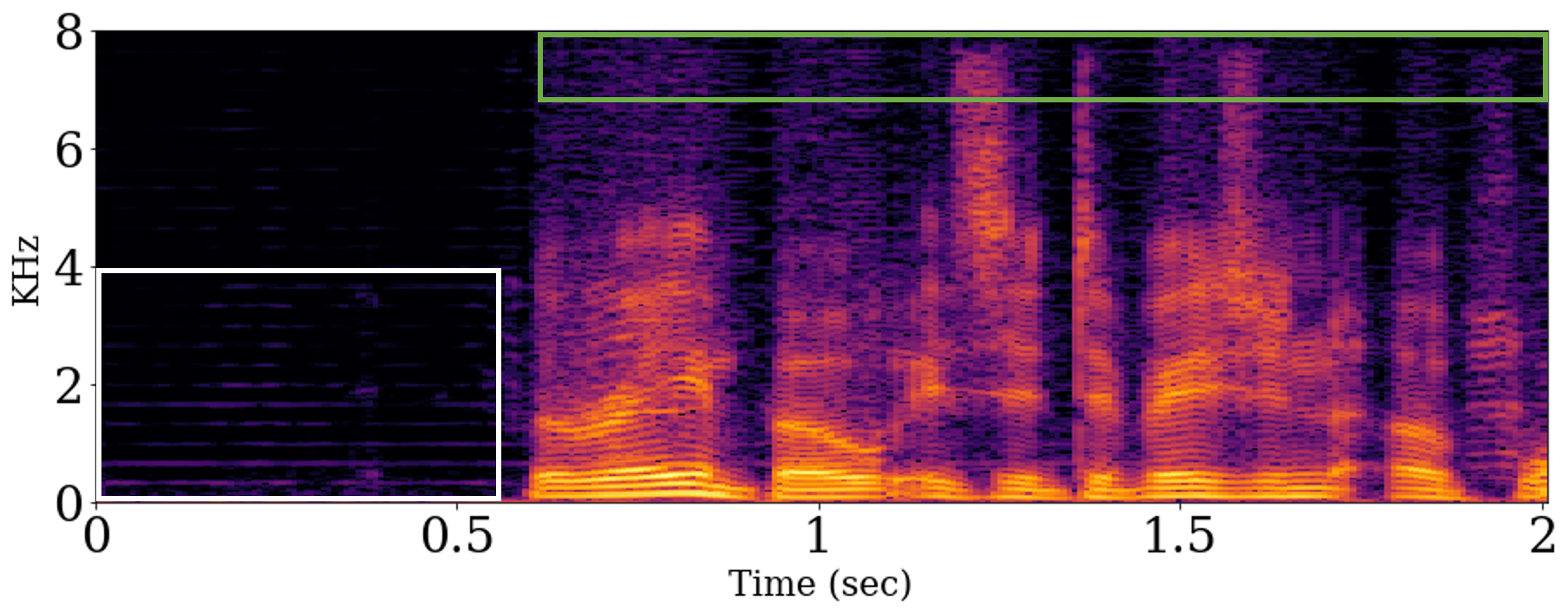}}
  \vspace{-0.1cm}
  \centerline{(g) WaveCRN*}\medskip
\end{minipage}
\hfill
\begin{minipage}[b]{0.49\linewidth}
  \centering
  \vspace{-0.2cm}
  \centerline{\includegraphics[width=\textwidth]{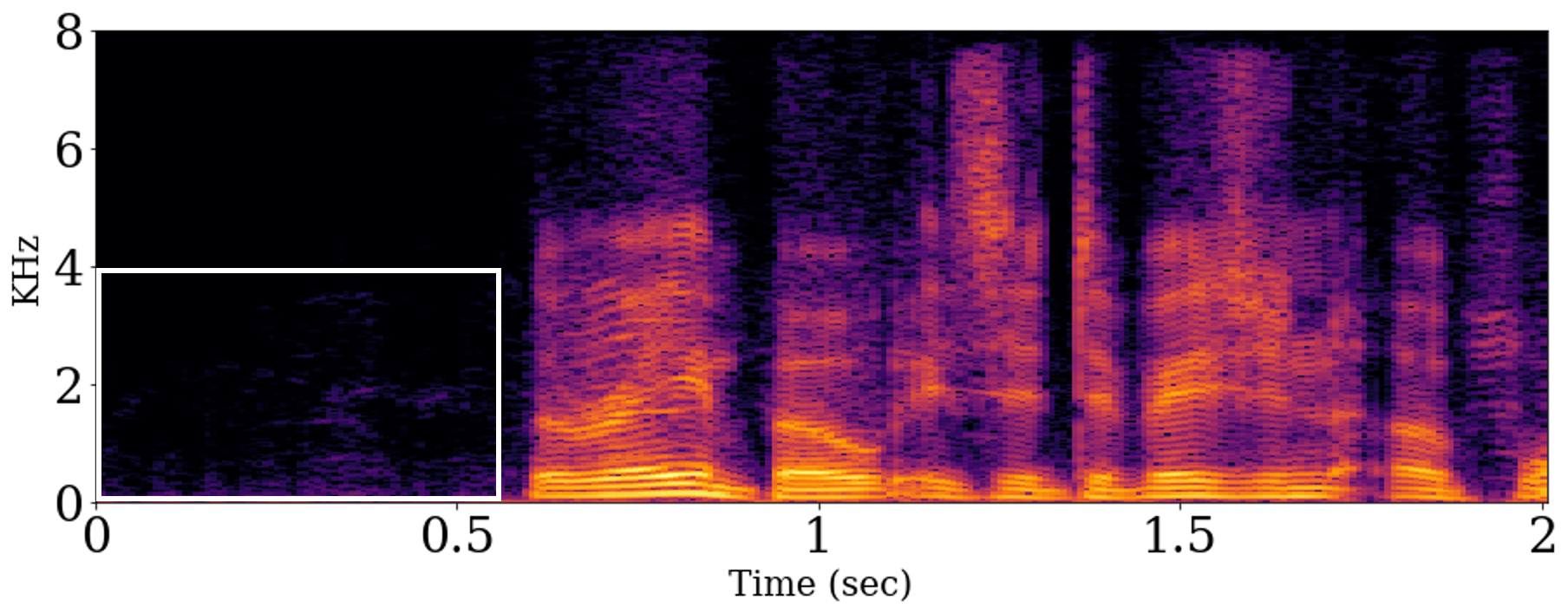}}
  \vspace{-0.1cm}
  \centerline{(h) WaveCRN}\medskip
\end{minipage}
\vspace{-0.9cm}
\caption{Magnitude spectrograms of noisy, clean and enhanced speech by LPS--SRU, LPS--SRU*, WaveCBLSTM, WaveCBLSTM*, WaveCRN, and WaveCRN*, where models marked with * directly generate enhanced speech without using RFM. \textcolor{black}{The improvements of WaveCRN over other methods are highlighted with green (high-frequency parts) and white blocks (silence).}}
\vspace{-0.5cm}
\label{fig:denoise}
\end{figure}
%
In \cite{sru}, SRU has been proved to yield a performance comparable to that of LSTM, but with better parallelism.  \textcolor{black}{The dependency between gates in LSTM leads to slow training and inference.} \textcolor{black}{In contrast, all gates in SRU depend only on the input of the current time, and the sequential correlation is captured by adding highway connections between recurrent layers. Therefore, the gates in SRU are computed simultaneously. In the forward pass, the time complexity of SRU and LSTM are $O(T\cdot N\cdot C)$ and $O(T\cdot N\cdot C^2)$, respectively. The above-mentioned advantages make SRU appropriate to combine it with CNN. Some studies \cite{veit2016residual, de2020batch} depict ResNet as an ensemble of relatively shallow paths of its sub-nets. Since SRU has highway connections and recurs over time, it can be regarded as an ensemble for discrete modeling of dependency within a sub-sequence.}
\section{Experiments}
\label{sec:part3}
\subsection{Experimental Setup}
\label{ssec:subhead}
\subsubsection{Speech Denoising}
\label{ssec:subsubhead}
For the speech denoising task, an open-source dataset \cite{NDTSEA} was used, which combines the voice bank corpus \cite{VCTK} and the DEMAND corpus \cite{DEMAND}. \textcolor{black}{Similar to previous works \cite{SEGAN, waveunet, Rethage2017Wavenet, yao2019coarse, DFL, pascual2019time}, we downsampled the speech data to 16 kHz for training and testing.} In the voice bank corpus, 28 out of the 30 speakers were used for training, \textcolor{black}{and 2 speakers were used for testing}. For the training set, the clean speech was contaminated with 10 types of noise at 4 SNR levels (0, 5, 10, and 15 dB). For the testing set, the clean speech was contaminated with 5 types of unseen noise at the other 4 SNR levels (2.5, 7.5, 12.5, and 17.5 dB).
\subsubsection{Compressed (2-bit) Speech Restoration}
\label{ssec:subsubhead}
For the compressed speech restoration task, we used the TIMIT corpus \cite{TIMIT}. The original speech samples were recorded in a 16 kHz and 16-bit format. In this set of experiments, each sample was compressed into a 2-bit format (represented by -1, 0, or +1). In this way, we save 87.5\% of the bits, thereby reducing the data transmission and storage requirements. We believe that this compression scheme is potentially applicable to real-world internet of things (IoT) scenarios. Note that the same model architecture was used in both denoising and restoration tasks. \textcolor{black}{The +1, 0, or -1 value of each compressed sample was first mapped to a floating-point representation, and thus the waveform-domain SE system could be readily applied to restore the original uncompressed speech.}
Expressing the original speech as $\hat{y}$ and the compressed speech as $\mathrm{sgn}(\hat{y})$, the optimization process becomes
\textcolor{black}{
\begin{equation}
    \arg\min_{\theta} \lVert \hat{y} - g_{\theta}(\mathrm{sgn}(\hat{y})) \rVert_{1},
\end{equation}
}
where $g_{\theta}$ denotes the SE process.
\begin{table}[t]
\caption{Results of the speech denoising task. A higher score indicates better performance. Bold values indicate the best performance for a specific metric. Models marked with * directly generate enhanced speech without using RFM.}
\centering
\small
\begin{tabular}{l|c c c c c}
\hline
Model  & PESQ & CSIG & CBAK & COVL & SSNR\\
\hline
\hline
\textbf{Noisy} & 1.97 & 3.35 & 2.44 & 2.63 & 1.68\\
\textbf{Wiener} & 2.22 & 3.23 & 2.68 & 2.67 & 5.07\\
\textbf{SEGAN} \cite{SEGAN} & 2.16 & 3.48 & 2.94 & 2.80 & 7.73\\
\textbf{Wavenet} \cite{Rethage2017Wavenet} & - & 3.62 & 3.23 & 2.98 & -\\
\textbf{Wave-U-Net} \cite{waveunet} & 2.62 & 3.91 & 3.35 & 3.27 & 10.05\\
\hline
\hline
\textbf{LPS--SRU*}& 2.21 & 3.28 & 2.86 & 2.73 & 6.18\\
\textbf{WaveCBLSTM*}& 2.39 & 3.19 & 3.08 & 2.76 & 8.78\\
\textbf{WaveCRN*}& 2.46 & 3.43 & 3.04 & 2.89 & 8.43\\
\textbf{LPS--SRU}& 2.49 & 3.73 & 3.20 & 3.10 & 9.17\\
\textbf{WaveCBLSTM}& 2.54 & 3.83 & 3.25 & 3.18 & 9.33\\
\textbf{WaveCRN}& \textbf{2.64} & \textbf{3.94} & \textbf{3.37} & \textbf{3.29} & \textbf{10.26}\\
\hline
\end{tabular}
\vspace{-0.3cm}
\label{tab:denoise}
\end{table}
\subsubsection{Model Architecture}
\label{ssec:subsubhead}
In the input module, \textcolor{black}{the number of channels, kernel size, and stride size was set to 256, 0.006 s, and 0.003 s, respectively}. The input audio was padded to make it divisible by the stride size. The size of the hidden state of Bi-SRU was set to the number of channels (with 6 stacks). Next, all the hidden states were linearly mapped to half dimension to form a mask and element-wisely multiplied by the feature map. Finally, in the waveform generation step, a transposed convolutional layer was used to map the 2D feature map into a 1D sequence, which was passed through a hyperbolic tangent activation function to generate the predicted waveform. \textcolor{black}{The $l_1$ norm was used as the objective function for training WaveCRN. For a fairer comparison of the model architectures, we mainly compare WaveCRN with other SE systems also trained using the $l_1$ norm.}


%

\subsection{Experimental Results}
\label{ssec:subhead}
\subsubsection{Speech Denoising}
\label{ssec:subsubhead}
For the speech denoising task, we used five evaluation metrics from \cite{metric}: \textbf{CSIG} (signal distortion), \textbf{CBAK} (background intrusiveness), \textbf{COVL} (overall quality using the scale of the mean opinion score) and \textbf{PESQ} that reveal the speech quality, and \textbf{SSNR} (segmental signal-to-noise ratio). Table \ref{tab:denoise} presents the results. The proposed model was compared with Wiener filtering, SEGAN, two well-known SE models that use the same $l_1$ loss (i.e., Wavenet and Wave-U-Net), \textcolor{black}{LPS--SRU that uses the LPS feature as input}, and WaveCBLSTM that combines CNN and BLSTM. LPS--SRU was implemented by replacing the 1D convolutional input module and the transposed 1D convolutional output module in Fig. \ref{fig:model} with STFT and inverse STFT modules.
WaveCBLSTM was implemented by replacing the SRU in Fig. \ref{fig:model} with LSTM. The combination of CNN and LSTM for processing speech signals has been widely investigated \cite{CRN1, CRN2, li2019single}. In this study, we aim to show that SRU is superior to LSTM in terms of the denoising capability and computational efficiency, when applied to waveform-based SE. \textcolor{black}{As can be clearly seen from Table \ref{tab:denoise}, WaveCRN outperforms other models in terms of all perceptual and signal-level evaluation metrics.} 

\textcolor{black}{We first investigated the effect of RFM. As shown in Table \ref{tab:denoise}, LPS--SRU, WaveCBLSTM, and WaveCRN are better than their counterparts without RFM (LPS--SRU*, WaveCBLSTM*, and WaveCRN*). Notably, unlike WaveCBLSTM and WaveCRN that use waveforms as input, LPS--SRU enhances the audio in the spectral domain. Fig. \ref{fig:denoise} shows the magnitude spectrograms of noisy, clean, and enhanced speech utterances. Two observations can be drawn from the figure. First, RFM notably eliminates noise components in the high-frequency region (green blocks) and silence parts (white blocks). This observation is consistent with the results in Table \ref{tab:denoise}: the models with RFM achieve higher SSNR scores and speech quality. Second, as shown in Fig. \ref{fig:denoise}(e), without RFM, the high-frequency region cannot be completely restored. Comparing Fig. \ref{fig:denoise}(e) and Fig. \ref{fig:denoise}(g), WaveCBLSTM* has a cleaner estimation than WaveCRN* in the silence parts, but the loss of the high-frequency region deteriorates the audio quality, which can be found in Table \ref{tab:denoise}. Compared with WaveCRN*, WaveCBLSTM* has a higher CBAK score but lower PESQ and SSNR scores.} \textcolor{black}{Next, Table \ref{tab:params} presents a comparison of the execution time and parameters of the WaveCRN and WaveCBLSTM. Under the same hyper-parameter settings (number of layers, dimension of hidden states, number of channels, etc.), the training process of WaveCRN is 15.45 ($(38.1 + 59.86) / (2.07 + 4.27)$) times faster than that of WaveCBLSTM, and the number of parameters is only 51\%. The forward pass is 18.41 times faster, which means 18.41 times faster in inference.}
\begin{figure}[t]
\begin{minipage}[b]{0.49\linewidth}
  \centering
  \centerline{\includegraphics[width=\textwidth]{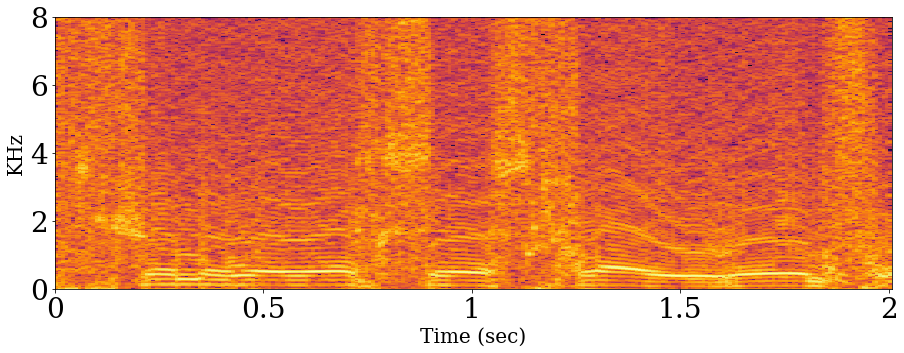}}
  \vspace{-0.1cm}
  \centerline{(a) Compressed}\medskip
\end{minipage}
\begin{minipage}[b]{0.49\linewidth}
  \centering
  \centerline{\includegraphics[width=\textwidth]{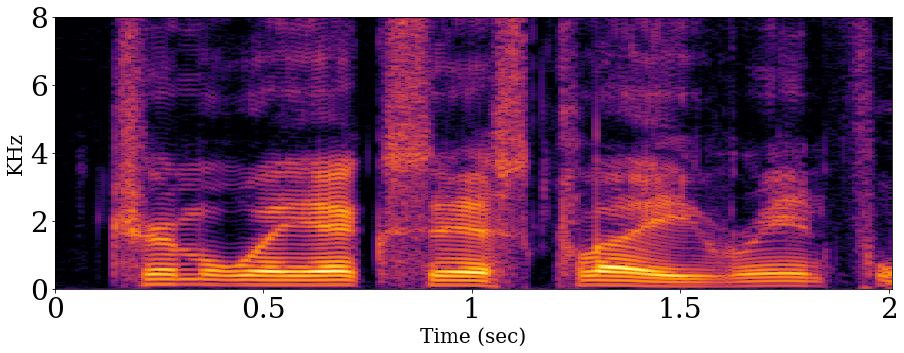}}
  \vspace{-0.1cm}
  \centerline{(b) Ground Truth}\medskip
\end{minipage}
\begin{minipage}[b]{0.49\linewidth}
  \centering
  \vspace{-0.2cm}
  \centerline{\includegraphics[width=\textwidth]{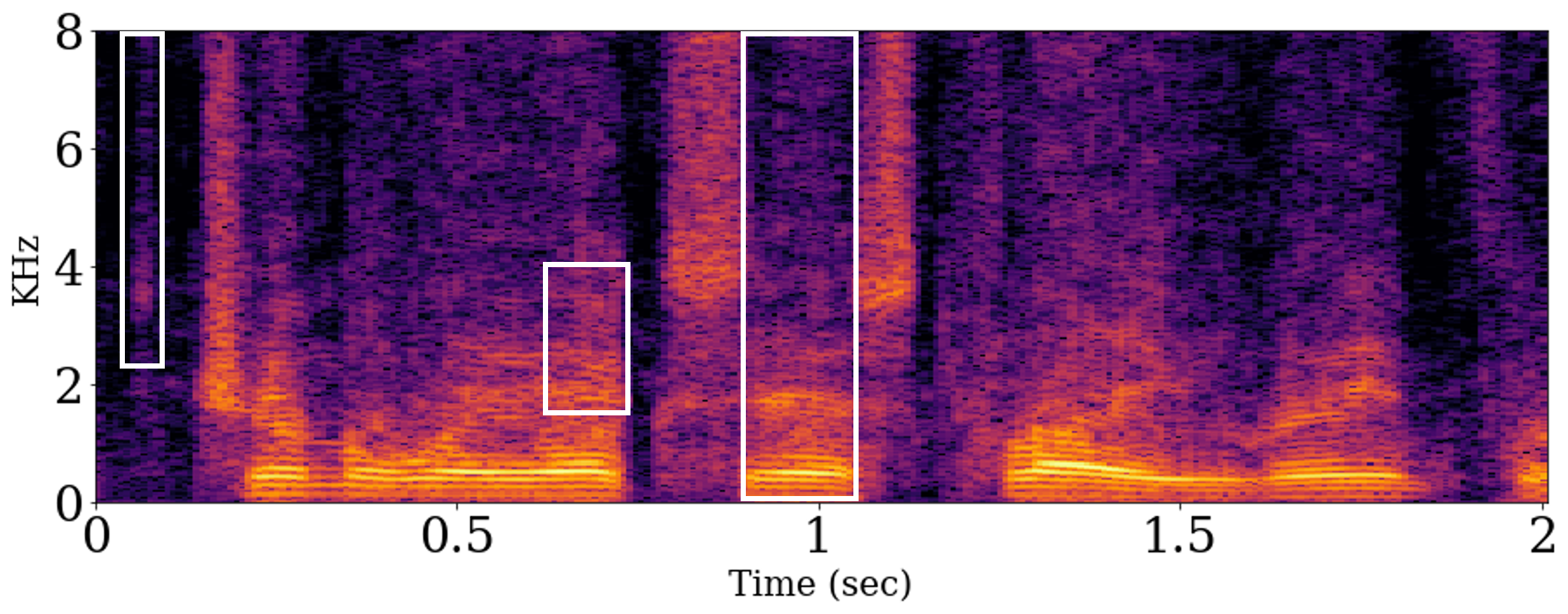}}
  \vspace{-0.1cm}
  \centerline{(c) LPS--SRU}\medskip
\end{minipage}
\begin{minipage}[b]{0.49\linewidth}
  \centering
  \vspace{-0.2cm}
  \centerline{\includegraphics[width=\textwidth]{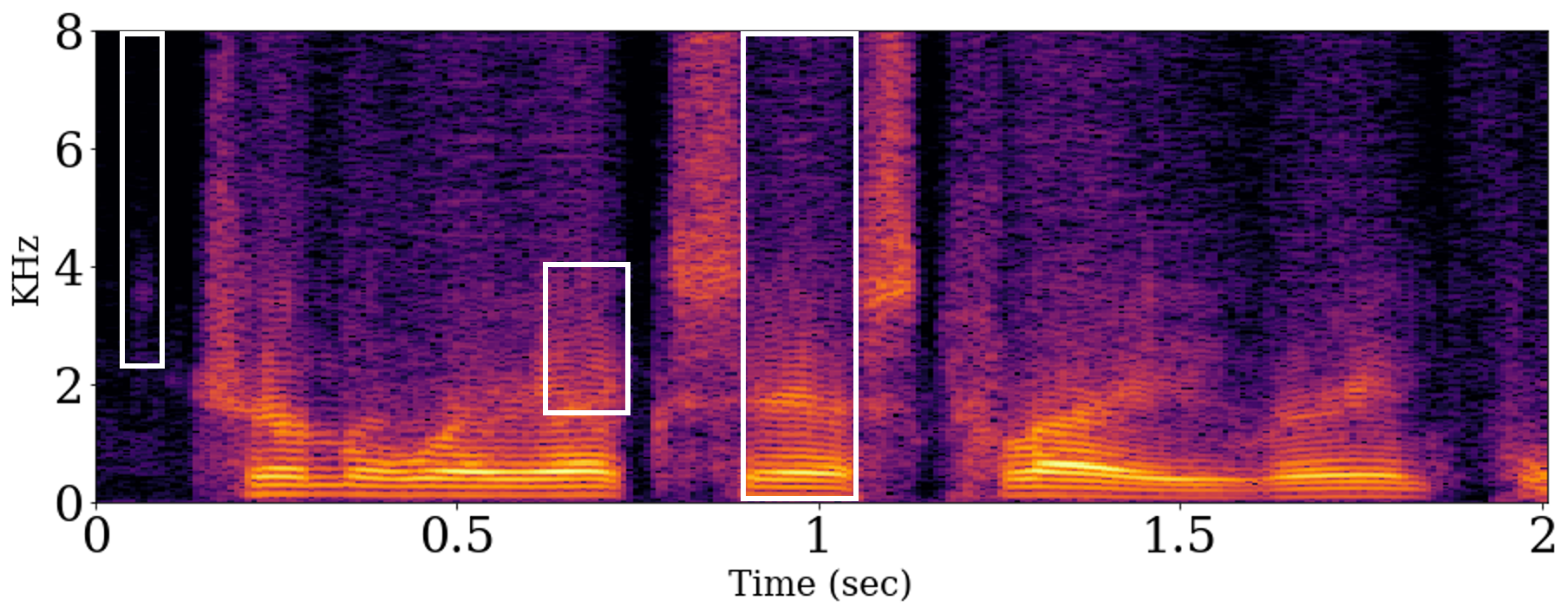}}
  \vspace{-0.1cm}
  \centerline{(d) WaveCRN}\medskip
\end{minipage}
\vspace{-0.5cm}
\caption{Magnitude spectrograms of original, compressed, and restored speech by LPS--SRU and WaveCRN.}
\vspace{-0.5cm}
\label{fig:restoration}
\end{figure}
\begin{table}[t]
\caption{A comparison of execution time and number of parameters of WaveCRN and WaveCBLSTM with same hyper-parameters. \textcolor{black}{This experiment was performed in an environment setting that used a 48-core CPU at 2.20 GHz and a Titan Xp GPU with 12 GB VRAM.} The first row and the second row show the execution time of the forward and the back-propagation passes for a 1-second waveform input in a batch of 16, and the third row presents the number of parameters.}
\centering
\small
\begin{tabular}{l|c c}
\hline
Model  & \textbf{WaveCBLSTM} & \textbf{WaveCRN}\\
\hline
\hline
Forward ($10^{-3}$sec) & 38.1 $\pm$ 2.1 & \textbf{2.07 $\pm$ 0.07}\\
Back-propagation ($10^{-3}$sec) & 59.86 $\pm$ 1.13 & \textbf{4.27 $\pm$ 0.09}\\
\#parameters (K) & 9093 & \textbf{4655}\\
\hline
\end{tabular}
\label{tab:params}
\vspace{-0.5cm}
\end{table}
\subsubsection{Compressed Speech Restoration}
\label{ssec:subsubhead}
\begin{figure}[t]
\begin{minipage}[b]{\linewidth}
  \centering
  \centerline{\includegraphics[width=0.49\textwidth]{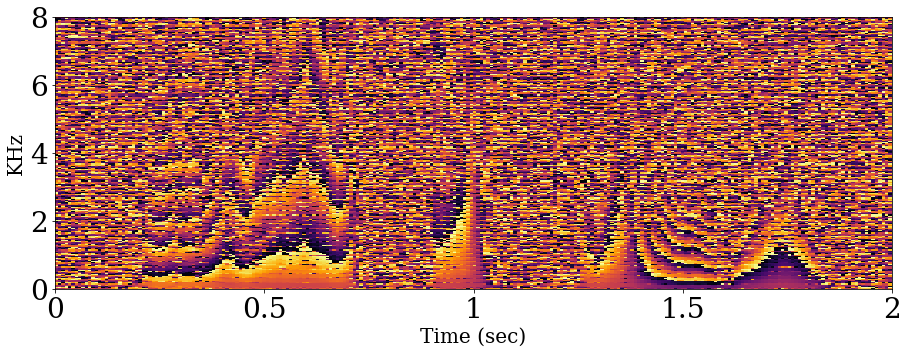}}
  \vspace{-0.1cm}
  \centerline{(a) Ground Truth}\medskip
\end{minipage}
%
%
\begin{minipage}[b]{0.49\linewidth}
  \centering
  \vspace{-0.2cm}
  \centerline{\includegraphics[width=\textwidth]{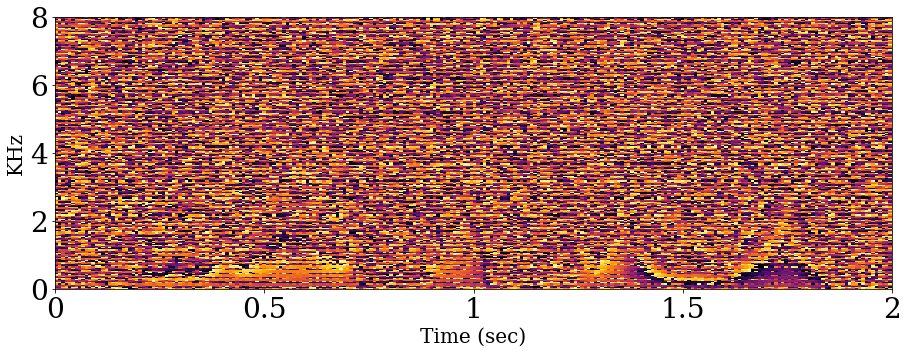}}
  \vspace{-0.1cm}
  \centerline{(b) LPS--SRU}\medskip
\end{minipage}
\begin{minipage}[b]{0.49\linewidth}
  \centering
  \vspace{-0.2cm}
  \centerline{\includegraphics[width=\textwidth]{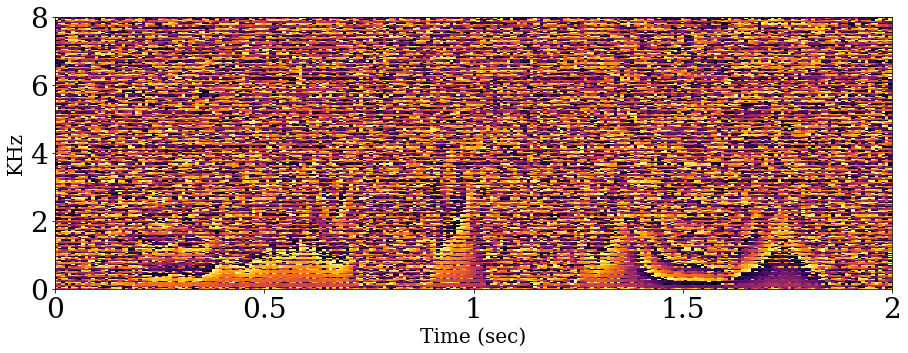}}
  \vspace{-0.1cm}
  \centerline{(c) WaveCRN}\medskip
\end{minipage}
\vspace{-0.5cm}
\caption{Instantaneous frequency spectrograms of uncompressed and restored speech by LPS--SRU and WaveCRN.}
\vspace{-0.5cm}
\label{fig:instafreq}
\end{figure}

For the compressed speech restoration task, we applied WaveCRN and LPS--SRU to transform the compressed speech to the uncompressed one. In LPS--SRU, the SRU structure was identical to that used in WaveCRN, but the input was the LPS, and the STFT and inverse STFT were used for speech analysis and reconstruction, respectively. The performance was evaluated in terms of the PESQ and STOI scores. From Table \ref{tab:restoration}, we can see that WaveCRN and LPS--SRU improve the PESQ score from 1.39 to 2.41 and 1.97, and the STOI score from 0.49 to 0.86 and 0.79. Both the \textcolor{black}{approaches} achieve significant improvements, while WaveCRN clearly outperforms LPS--SRU.
We can observe from Fig. \ref{fig:restoration}(a) and \ref{fig:restoration}(b) that the speech quality is notably reduced \textcolor{black}{in 2-bit format}, especially in the silence part and the high-frequency region. However, the spectrograms of speech restored by WaveCRN and LPS--SRU present a clearer structure, as shown in Fig. \ref{fig:restoration}(c) and \ref{fig:restoration}(d). In addition, the white-block regions show that WaveCRN can restore speech patterns more effectively than LPS--SRU. \textcolor{black}{Fig. \ref{fig:instafreq} shows the instantaneous frequency spectrograms. As expected, the LPS--SRU recovers the waveform through the compressed phase spectrogram; hence, WaveCRN preserves more details 
of the phase spectrogram by directly using the waveform as input without losing phase information.}
%
\begin{table}[t]
\caption{The results of the compressed speech restoration task.}
\centering
\begin{tabular}{l|c c }
\hline
 Model & PESQ & STOI\\
\hline
\hline
\textbf{Compressed} & 1.39 & 0.49 \\
\hline
\textbf{LPS--SRU} & 1.97 & 0.79 \\
\textbf{WaveCRN} & \textbf{2.41} & \textbf{0.86} \\
\hline
\end{tabular}
\label{tab:restoration}
\vspace{-0.5cm}
\end{table}
\section{Conclusions}
\label{sec:part4}
This paper proposed the WaveCRN E2E SE model. WaveCRN uses a bi-directional architecture to model the sequential correlation of extracted features. The experimental results show that WaveCRN achieves outstanding denoising capability and computational efficiency compared with related works using $l_1$ loss. The contributions of this study are fourfold: (a) WaveCRN is the first work that combines SRU and CNN to perform E2E SE; (b) a novel RFM approach was derived to directly transform the noisy features to enhanced ones; (c) the SRU model is relatively simple yet yield comparable performance to other up-to-date SE models that use the same $l_1$ loss; (d) a new practical application (compressed speech restoration) was designed and its performance was tested; WaveCRN obtained promising results on this task. \textcolor{black}{This study focused on comparing the SE model architecture with the conventional $l_1$ norm loss. Our future work will explore adopting alternative perceptual and adversarial losses in the WaveCRN system.}

\vfill\pagebreak
\label{sec:refs}
\bibliographystyle{IEEEbib}
\bibliography{refs}
\end{document}